\documentclass[twocolumn,letterpaper,showpacs,aps,prl,floatfix,10pt]{revtex4-2}

\usepackage{graphicx}
\usepackage{verbatim}
\usepackage{dcolumn}
\usepackage{amsmath}
\usepackage{epsfig}
\usepackage{subfigure}
\usepackage{multirow}
\usepackage{siunitx}
\usepackage{graphicx}

\input{babarsym.tex}

\newcommand{\BaBarYear}      {24}
\newcommand{\BaBarNumber}    {001}
\newcommand{\BaBarType}      {PUB}  
\newcommand{\SLACPubNumber}  {241025}

\def\Sig           {\ensuremath{B^+ \to \Lambda_c^+ \, \psi_D}\xspace}
\def\LambdaCtopkpi {\ensuremath{\Lambda_c^+ \to p K^- \pi^+}\xspace}
\def\SigPsi        {\ensuremath{\psi_D}\xspace}
\def\SigLamb       {\ensuremath{\Lambda_c^+}\xspace}
\def\BtoLambdaPsi  {\ensuremath{B^0 \to \Lambda \, \psi_D }\xspace}
\def\BtoPPsi       {\ensuremath{B^+ \to p \,  \psi_D }\xspace}
\def\SigBF         {\ensuremath{{\rm BF}(B^+ \to \Lambda_c^+ \, \psi_D})\xspace}
\def\MPsi          {\ensuremath{m_{\psi_D}}\xspace}
\def\Mpkpi         {\ensuremath{m_{p K^+ \pi^-}}\xspace}
\def\Btag          {\ensuremath{B_{\rm tag}}\xspace}
\def\Bsig          {\ensuremath{B_{\rm sig}}\xspace}
\def\Lumi          {\ensuremath{431.0 \rm{~fb}^{-1}}\xspace}
\def\LumiRun3      {\ensuremath{32.5 \rm{~fb}^{-1}}\xspace}
\def\LumiSig       {\ensuremath{398.5 \rm{~fb}^{-1}}\xspace}
\def\sigmaPsi      {\ensuremath{\sigma_{\psi_D}}\xspace}
\def\SigEff        {\ensuremath{\epsilon_{\rm sig}}\xspace}

\def\bfLamdaCtopkpi{\ensuremath{ (6.24 \pm 0.28})\% \xspace}
\def\fBB    {\ensuremath{ 0.86 \pm 0.04 }\xspace}
\def\fqq    {\ensuremath{ 0.86 \pm 0.02 }\xspace}
\def\effmax {\ensuremath{6.6\times 10^{-5}\xspace}}
\def\effmin {\ensuremath{5.9\times 10^{-5}\xspace}}
\def\bgtot  {\ensuremath{0.4}\xspace}

\newcommand{\GeV}{~\mbox{GeV}}
\newcommand{\MeV}{~\mbox{MeV}}

\begin{document}

\begin{flushleft}
\babar-\BaBarType-\BaBarYear/\BaBarNumber \\ 
SLAC-PUB-\SLACPubNumber\\
\end{flushleft}

\title{ \large \bf\boldmath 
Search for baryogenesis and dark matter in \ensuremath{B^+ \to \Lambda_c^+  + {\rm invisible}} decays
}

\author{J.~P.~Lees}
\author{V.~Poireau}
\author{V.~Tisserand}
\author{E.~Grauges}
\author{A.~Palano}
\author{G.~Eigen}
\author{D.~N.~Brown}
\author{Yu.~G.~Kolomensky}
\author{M.~Fritsch}
\author{H.~Koch}\thanks{Deceased}
% ...
\collaboration{The \babar\ Collaboration}
\noaffiliation

\begin{abstract}
A mechanism of baryogenesis and dark matter production via $B$ meson oscillations and decays has recently been proposed to explain the observed dark matter abundance and matter-antimatter asymmetry in the universe. This mechanism introduces a low-mass dark sector particle ($\psi_D$) with a nonzero baryonic charge. We present a search for this new state in \Sig decays using data collected at the \FourS resonance by the \babar\ detector at SLAC, corresponding to an integrated luminosity of \Lumi. The search leverages the full reconstruction of the $\Bm$ meson in $\FourS \to \BpBm$ decays, accompanied by the reconstruction of a \SigLamb, to infer the presence of \SigPsi. No significant signal is observed, and limits on the \Sig branching fraction are set at the level of $1.6 \times 10^{-4}$ at 90\% confidence level for $0.94 < \MPsi < 2.99 \GeV$. These results set strong constraints on the parameter space allowed by $B$-meson baryogenesis.
\end{abstract}

\maketitle

The nature of dark matter (DM) and the baryon asymmetry of the universe (BAU) are among the deepest mysteries of modern physics. Recently, the $B$ mesogenesis  mechanism has been proposed to simultaneously generate the DM abundance and BAU~\cite{Elor:2018twp,Alonso-Alvarez:2021qfd}. This scenario postulates the existence of a dark sector particle with a nonzero baryonic charge, \SigPsi, and a heavy colored scalar $Y$ mediating quark-DM interactions. Out-of-thermal-equilibrium production of \bbbar pairs, followed by $C\!P$-violating \BzBzb oscillations and subsequent decay into a baryon and dark sector antibaryon  (with any number of additional mesons), results in a net conservation of baryon number, but with the observed BAU balanced by an equivalent dark matter antibaryon asymmetry. 

According to the $B$ mesogenesis mechanism, the BAU depends on the electric charge asymmetry in semileptonic $B$ meson decays and the total $B \rightarrow \mathcal{B} \, \mathcal{M} \, \psi_D $ branching fraction, where $\mathcal{B}$ denotes any type of baryon and $\mathcal{M}$ any number of additional mesons. Current bounds on the semileptonic asymmetries require a branching fraction ${\rm BF}(B \rightarrow \mathcal{B} \,  \mathcal{M} \,  \psi_D ) \gtrsim 10^{-4}$ for successful $B$ mesogenesis~\cite{Elor:2018twp}. The type of baryon depends on the operator mediating the quark-DM interaction, and several final states must be investigated to fully test this mechanism. Constraints on exclusive $B \rightarrow \mathcal{B} \,  \psi_D$ decays are derived using phase-space arguments~\cite{Alonso-Alvarez:2021qfd} or QCD light cone sum rules calculations~\cite{Elor:2022jxy} for the lightest baryons. In the case of $B \rightarrow \Lambda_c^+ \,  \psi_D$, the ratio of exclusive-to-inclusive decays varies between 10\% and 100\%, depending on the $\psi_D$ mass. Additionally, proton and DM stability require $0.94 \GeV < m_{\psi_{D}} < (m_{B} - m_{\mathcal{B}})$. A similar process involving charged $B^\pm$ mesons has also been proposed~\cite{Elahi:2021jia}, which could be partially constrained with the reaction studied hereafter.

\begin{figure}[ht]
\begin{center}
 \includegraphics[width=0.4\textwidth]{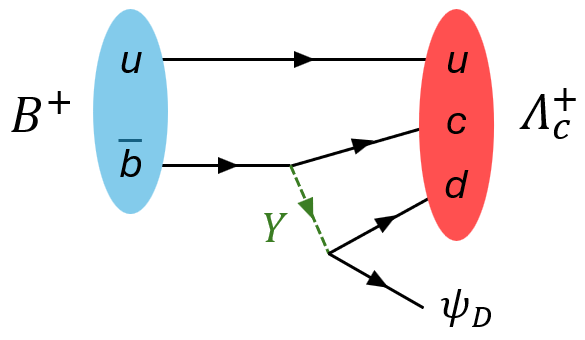}
\caption{Diagram of the decay of a $B$ meson into a $\Lambda_c^+$ and a dark sector antibaryon $\SigPsi$. The quark-DM interaction is mediated by a heavy scalar mediator $Y$. }
\label{figfeyn}
\end{center}
\end{figure}

Previous experimental searches by \babar\ and Belle have placed limits on the decay \BtoLambdaPsi~\cite{Belle:2021gmc,BaBar:2023rer}, and a recent measurement by \babar\ set bounds on \BtoPPsi~\cite{BaBar:2023dtq}. These results constrain operators $\mathcal{O}_{us}$ and $\mathcal{O}_{ud}$ describing $b \to \SigPsi u s$ and $b \to \SigPsi u d$ effective interactions, respectively~\cite{Elor:2018twp,Alonso-Alvarez:2021qfd}. The decay under study in this work, \Sig, probes an effective operator $\mathcal{O}_{cd}$ describing $b \to \SigPsi c d$ transitions, as shown in Fig.~\ref{figfeyn}. At present, only indirect constraints on the branching fraction of this final state have been derived at the level of $6 \times 10^{-4}$ at 90\% confidence level (CL)~\cite{Alonso-Alvarez:2021qfd}. Signature of $B$ mesogenesis could also be explored at the LHC experiments~\cite{Alonso-Alvarez:2021qfd, Rodriguez:2021urv, Kalsi:2024hfe} and at large volume neutrino experiments~\cite{Berger:2023ccd}.

We present herein a search for the decay \Sig~\footnote{charge conjugate modes are implied throughout this Letter.} based on the full \babar\ data set collected at the \FourS resonance. Because the final state includes missing energy from the undetected \SigPsi, we utilize a technique in which the accompanying \Bm meson, referred to as \Btag, is exclusively reconstructed in one of many hadronic decay modes. The remainder of the event, referred to as the \Bsig, is then examined for evidence of a signal decay, in the form of a reconstructed \SigLamb ($\to p K^- \pi^+$), accompanied by missing energy consistent with a \SigPsi candidate of mass \MPsi.

The \babar\ detector operated at the PEP-II asymmetric $e^+e^-$ storage
rings at SLAC between 1999 and 2008. The apparatus consists of concentric layers of detector subsystems arranged in a cylindrical geometry around the $e^+e^-$ interaction point. It is described in detail in Refs~\cite{BaBar:2001yhh, BaBar:2013byz}. Charged particle tracks are reconstructed using a five-layer double-sided silicon vertex tracker and a multiwire drift chamber with 40 radial layers. Momenta are determined based on track curvature measured with the tracking system within the uniform 1.5~T magnetic field of a superconducting solenoid oriented parallel to the collision axis. Protons, kaons, and pions are distinguished based on specific energy loss, $dE / dx$, in the tracking detectors, and measurements of Cherenkov photons emitted in an internally reflecting ring-imaging detector. Electromagnetic showers from photons and electrons are reconstructed in a CsI(Tl) electromagnetic calorimeter (EMC), consisting of a barrel region and a forward endcap, situated inside of the solenoid. Muons are identified based on their interactions in an instrumented magnetic flux return located outside of the solenoid. These measurements are combined to provide standardized particle identification (PID) selectors for various efficiencies and misidentification levels for common charged particle species: $e, \mu, \pi, K, p$. 

The analysis is based on the full \babar\ data set corresponding to a total integrated luminosity of \Lumi~\cite{BaBar:2013agn}, but about 5\% of the data sample is used to optimize the analysis and is subsequently discarded. The remaining data (\LumiSig) were not examined until the analysis procedure had been finalized. 

Signal Monte Carlo (MC) data sets for \Sig decays with subsequent \LambdaCtopkpi decays are generated with EvtGen~\cite{Lange:2001uf} based on a phase space model. Distinct MC samples are produced for \SigPsi masses of 1.0, 1.5, 2.0, 2.3, 2.5, 2.7, and $2.9 \GeV$ \footnote{Natural units, $\hbar = c = 1$ are used throughout this Letter.}. Backgrounds are studied using large samples of generic MC events equivalent to approximately 4--7 times the data luminosity. Samples of \BpBm and \BzBzb events are generated using EvtGen, while \ccbar and \qqbar ($q = u,d,s)$ are simulated using JETSET~\cite{Bierlich:2022pfr}. As the tau-pair background, simulated using KK~\cite{Jadach:1999vf}, is found to be negligible, its contribution was not included in the analysis. PHOTOS~\cite{Barberio:1993qi} is used to generate final-state radiation. Detector response is simulated using GEANT4~\cite{GEANT4:2002zbu}. 

The \Btag identification utilizes the hadronic $B$ reconstruction methodology described in \cite{BaBar:2014omp}. The \Btag candidates are reconstructed in $B \to SX$ by combining a ``seed'' ($S$) of  $ D^{(*)0}$, $D^{(*)\pm}$, $D_s^{(*)\pm}$ or  $J/\psi$,  with a hadronic system $X$ comprised of up to five charged or neutral kaons or pions, with a total charge of $\pm 1$ or $0$. The $D$ meson seeds are reconstructed via:
$D^+\to\KS\pip$, $\KS\pip\piz$, $\KS\pip\pim\pip$, $ K^-\pip\pip$, $K^-\pip\pip\piz$, $K^{+}\Km\pip$, $K^{+}\Km\pip\piz$; 
$\Dz\to K^-\pip$, $K^-\pip\piz$, $K^-\pip\pim\pip$, $\KS\pip\pim$, $\KS\pip\pim\piz$, $K^{+}\Km$, $\pip\pim$, $\pip\pim\piz$, $\KS\piz$; 
$D^{*+}\to \Dz\pip$, $\Dp\piz$; $\Dstarz\to \Dz\piz$, $\Dz\g$; $D_s^{*+}\to D_s^+\g$; $D_s^+\to\phi\pip$, and $\KS K^{+}$.  The $J/\psi$ seeds are reconstructed via $J/\psi \to e^+e^-$ and $\mu^+\mu^-$. Light neutral mesons are reconstructed as $\pi^0 \to \gamma \gamma$, $\KS \to  \pip\pim$, and $\phi \to  K^{+}K^{-}$. Seed candidates and any intermediate states are required to lie within a mass window around the expected particle mass. A fit including vertex and particle mass constraints to the seed and its daughters is performed.

Correct combinations of $S$ and $X$ are then determined by requiring that the resulting \Btag candidate be consistent with $B$ meson kinematics based on the two variables $\mes \equiv \sqrt{(E^*_{\rm CM}/2)^2 - \Vec{p^*}^2_{B_{\rm tag}}}$ and $\DeltaE \equiv E^*_{B_{\rm tag}} - E^*_{\rm CM}/2 $, where $E^*_{\rm CM}/2$ is the center-of-mass frame beam energy, and $E^*_{B_{\rm tag}}$ and $\Vec{p^*}_{B_{\rm tag}}$ are the energy and three-momentum vector of the reconstructed \Btag candidate in the center-of-mass frame, respectively. Correctly reconstructed \Btag candidates have \DeltaE close to zero and \mes consistent with the nominal $B$ meson mass. Candidates are initially retained if $|\DeltaE| <0.12 \GeV$ and $\mes > 5.2 \GeV$. The latter requirement retains a sideband region below the $B$ mass for studies of combinatorial backgrounds. If more than one \Btag candidate is reconstructed in an event, the candidate in the highest-purity decay mode is retained, where the purity is defined as the ratio of correctly reconstructed \Btag candidates to total reconstructed candidates. If the purities of the two candidates are equal, then the one with the lowest $|\DeltaE|$ is retained. 

Once a single \Btag candidate has been selected, all remaining detector activity in the event is attributed to the \Bsig. Additional selection criteria are then applied in order to identify signal candidates, based on the \Btag, \Bsig, and the overall event properties. Only events possessing a charged \Btag (i.e., $\Bm$) along with exactly three high-quality charged tracks are selected, requiring that the total event charge be zero. High-quality tracks are defined as tracks having at least 12 hits in the drift chamber and a distance of closest approach to the interaction point less than 1.5 cm (2.5 cm) in the plane transverse (parallel) to the beam axis. Up to one additional low-quality charged track is permitted, usually arising from interactions with the detector material or reconstruction artifacts (its charge is not counted towards the total event charge). The \SigLamb candidate is reconstructed by combining the three high-quality charged tracks, one of which is required to satisfy loose PID criteria for a proton, and another one for a kaon. No PID requirements are imposed on the third track, assumed to be a $\pi^+$. A fit to the \SigLamb candidate is performed, including vertex constraints and appropriate mass hypotheses for the tracks, and a loose $\chi^2$ constraint is imposed to ensure convergence of the fit. Multiple \SigLamb candidates (from particle misidentification) are reconstructed in less than 1\% of the events. In such cases, all candidates are considered. 

In addition to the reconstructed \Btag and \SigLamb, both signal and background events typically contain additional detector activity in the form of reconstructed energy clusters in the EMC. For signal events, these can arise due to accelerator beam backgrounds, bremsstrahlung, and reconstruction artifacts. For background events they can also arise from $\pi^0$ decay daughters or other primary-interaction decay products. Consequently, the energy, multiplicity, and other properties of these additional clusters can be used to differentiate between signal and background events. Clusters are considered to be photon candidates if their reconstructed energies exceed $25 \MeV$. Neutral pion candidates are obtained by combining pairs of photons having an invariant mass $m_{\gamma \gamma}$ within $15 \MeV$ of the nominal $\pi^0$ mass.

At this stage of the selection, the background is dominated by continuum 
$e^+e^- \to q\bar{q} ~ (q = u,d,s,c)$ events, as well as by \BB events with an incorrectly reconstructed \Btag. The generic MC samples are known to overestimate the \Btag yield for both correctly reconstructed \Btag events and combinatorial backgrounds~\cite{BaBar:2013npw, BaBar:2016wgb, BaBar:2019awu}. This discrepancy arises from a combination of mismodeled branching fractions in the simulation, and multiple small discrepancies in charged and neutral particle reconstruction efficiencies compared with data. The observed discrepancy depends on the specific \Btag selection requirements that are imposed, and on which \Btag decay modes are retained. Correction factors $f_{B\bar{B}} = \fBB$ and $f_{q\bar{q}} = \fqq$ are obtained for \BB and continuum backgrounds, respectively, by simultaneously fitting several variables describing the global event shape and the \Btag meson. 

\begin{figure}[htb!]
\begin{center}
 \includegraphics[width=0.48\textwidth]{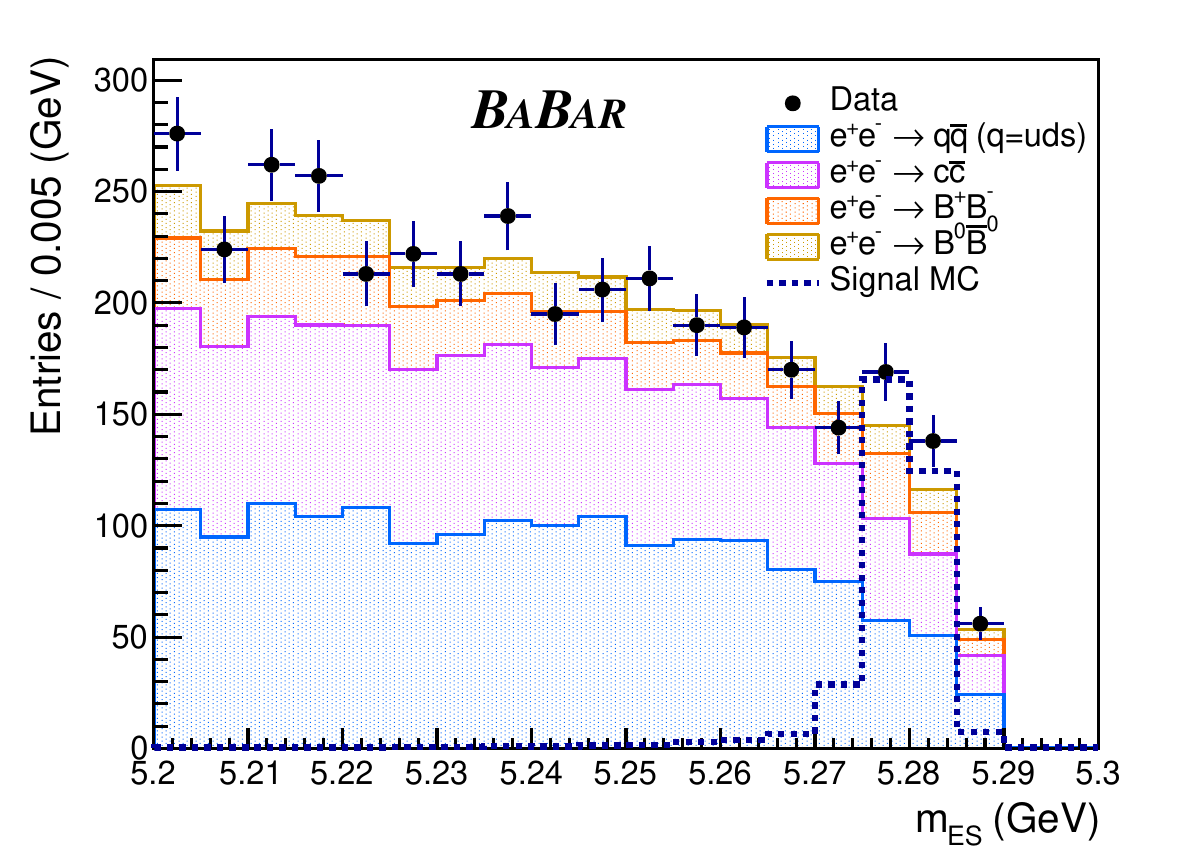}
\caption{The distribution of the $B$-tag energy-substituted mass ($m_{ES}$) for the data (black circles), together with the various background MC samples (stacked histograms). The signal MC (dashed line) is shown for $\MPsi = 2.0\GeV$ with arbitrary scaling. The uncertainties are purely statistical.}
\label{figmes}
\end{center}
\end{figure}

\begin{figure}[htb!]
\begin{center}
\includegraphics[width=0.48\textwidth]{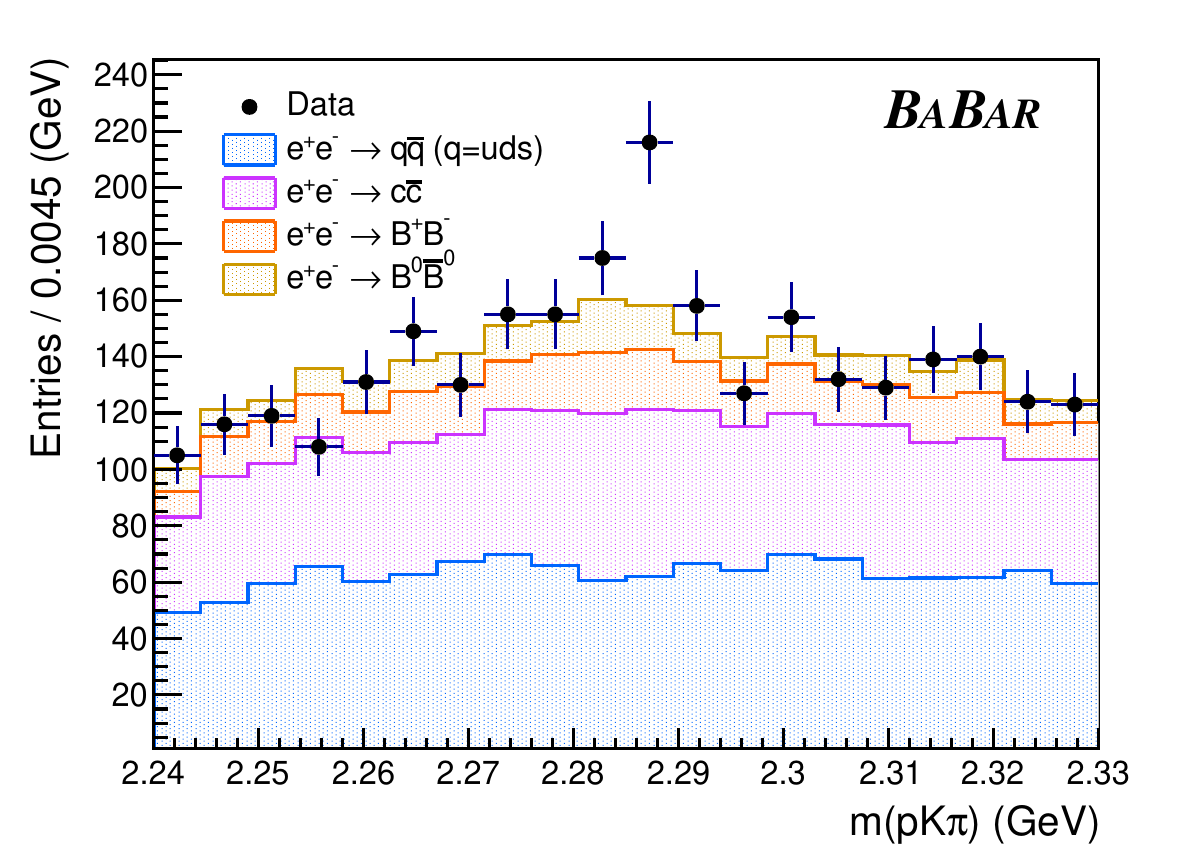}
\caption{The distribution of the $pK\pi$ mass ($m_{pK\pi}$) for the data (black circles) for the sideband region $m_{ES} < 5.27 \GeV$ together with the various background MC samples (stacked histograms). The uncertainties are purely statistical.}
\label{figmpkpi}
\end{center}
\end{figure}

Good agreement between data and MC is observed in all selection variables after the application of these correction factors. The uncertainty in the \BB correction factor is propagated as a systematic uncertainty in the signal efficiency.
Figures~\ref{figmes} and~\ref{figmpkpi} show the distribution of \mes and \Mpkpi for data and MC events after applying the correction factors. No obvious peaking structure is visible at the $B$ mass, but a clear \SigLamb peak is evident in data for the sideband region $m_{ES} < 5.27 \GeV$, which is not properly modeled in the background MC. Comparison of this peak in data to the signal MC provides a validation of the \SigLamb modeling in the signal MC, with the mass resolution and peak position found to be consistent within statistical uncertainties. This mismodeled continuum \SigLamb component is highly suppressed by the subsequent signal selection and has a negligible impact on the final results. 

\begin{figure}[htb]
\begin{center}
 \includegraphics[width=0.5\textwidth]{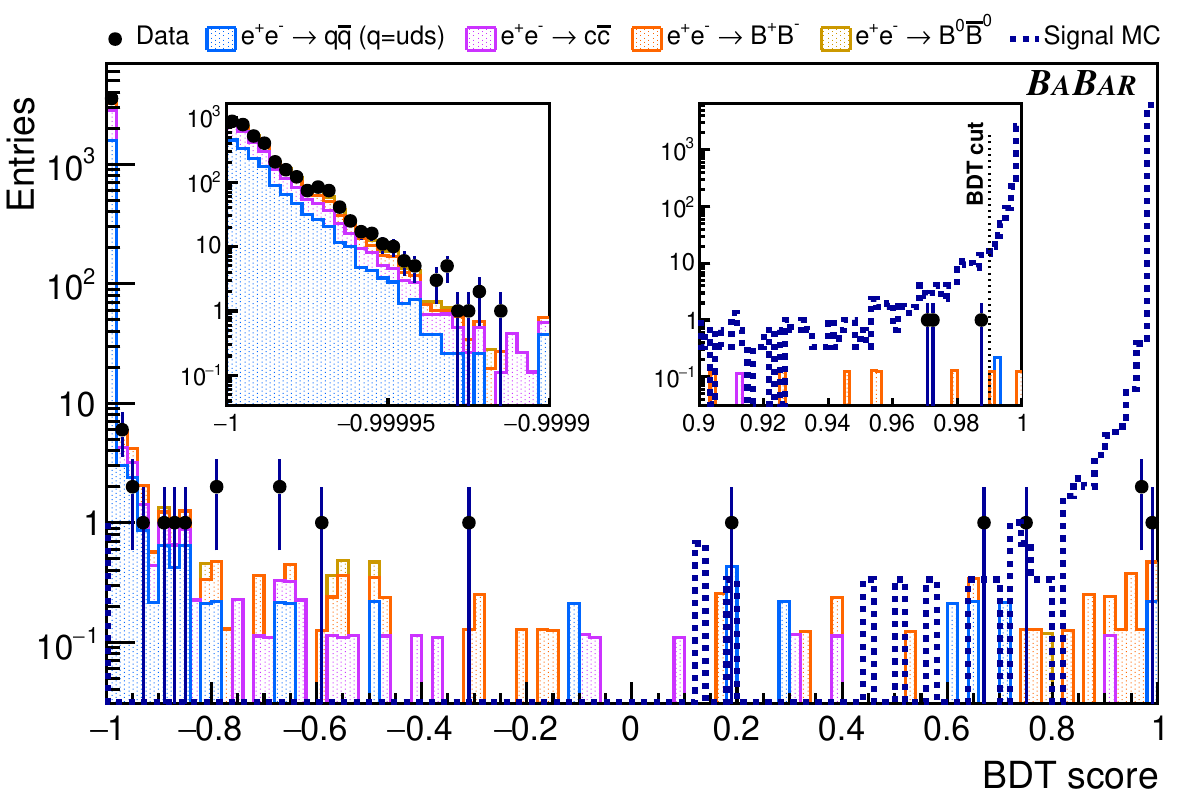}
\caption{The distribution of the BDT score for the unblinded data set together with the various background MC samples (stacked histogram). The signal MC (dotted line) is shown for $\MPsi = 2.0\GeV$ with arbitrary scaling. The uncertainties are purely statistical.}
\label{fig:BDT}
\end{center}
\end{figure}

To further reduce backgrounds, a second stage of signal selection is imposed, based on a multivariate classifier utilizing a boosted decision tree (BDT). The BDT inputs comprise global event-shape variables, measures of the \Btag and \SigLamb reconstruction quality, and quantities related to the event missing energy and additional neutral clusters. The 14 input variables are described in Table~\ref{tab:BDTinputs}. They are only weakly correlated, with the exception of the purity and integrated purity variables. The BDT is trained on signal MC samples, with \SigPsi masses spanning the range $1.0 < \MPsi < 2.9\GeV$, and background MC samples. A clear separation is obtained between signal and background events, as shown in Fig~\ref{fig:BDT}. 

\begin{table}[htb!]
    \centering
    \begin{tabular}{p{1.5cm} | p{1cm} | p{5.5cm}}
     Input   &  Type & Description  \\ \hline \hline
   $R_2$   &  Event shape & Ratio of the second-to-zeroth Fox-Wolfram moment~\cite{Fox:1978vw} computed using all tracks and neutral clusters  \\ 
    purity & \Btag & Fraction of correctly reconstructed $B$ mesons in the candidate \Btag mode \\
   integrated purity & \Btag & Overall purity of a $B$ tag sample containing all events having a tag purity equal or greater than that of the $B$ candidate\\ 
   $B_{\rm mode}$ & \Btag & Reconstructed decay mode of the \Btag \\ 
   \mes & \Btag & \Btag invariant mass    \\
   \DeltaE & \Btag &  Difference between the \Btag energy and the beam energy \\
   $B_{\rm thrust}$ & \Btag &  The magnitude of the \Btag thrust  \\
   $B_{\rm{thrust}Z}$ & \Btag & Component of the $B_{\rm thrust}$ along the $e^+e^-$ collision axis ($z$-axis)    \\ 
   \Mpkpi & \SigLamb & Reconstructed invariant mass of the $\SigLamb$ candidate  \\
   $\chi^2$ & \SigLamb & $\chi^2$ of the fit of the $\SigLamb$ candidate  \\ 
   $N_{\rm neut}$ & ECL  & Total number of additional neutral clusters \\
   $N_{\pi^0}$ & ECL & Number of additional $\pi^0$ candidates \\
   $E_{\rm extra}$ & ECL & Sum of the energies of all additional neutral clusters \\ 
   $\cos\theta_{\psi_D}$ & \SigPsi & Cosine of the polar angle of the missing energy 4-vector in the laboratory frame.
    \end{tabular}
    \caption{Inputs to the boosted decision tree classifier.}
    \label{tab:BDTinputs}
\end{table}

The BDT selection is optimized globally, rather than for an individual \MPsi mass hypothesis, using a figure of merit (FOM) similar to the Punzi FOM~\cite{Punzi:2003bu}. The result is a single requirement on the BDT score greater than 0.99. This criterion is found to be almost fully efficient for signal events surviving the preselection and \SigLamb reconstruction requirements. The MC and data yields show good consistency across the BDT score distribution below the signal region cut. Due to the low statistical precision in the MC samples, the distribution of background events in the signal region is modeled using simulated data in the wide range $[-0.8,1]$ in the BDT score. A total of \bgtot background events is predicted in the mass range $0.94 < \MPsi < 2.99 \GeV$. Unblinding the signal region yields no selected events in the full data set, consistent with the background expectation.

We set limits on the \Sig branching fraction as a function of \MPsi, determined from the $\psi_D$ four-momentum obtained from the information in the rest of the event. The \SigPsi four-momentum is derived by subtracting the \Btag and \SigLamb four-momenta from the precisely known center-of-mass four-momentum. The \MPsi resolution, \sigmaPsi, is determined from signal MC as a function of mass and interpolated between the simulated mass points. It decreases from approximately $60\MeV$ to $20\MeV$ across the mass range $0.94< \MPsi < 2.99 \GeV$.

The signal efficiency, \SigEff, decreases roughly linearly from $\effmax$ to $\effmin$ with increasing \MPsi.  The low efficiency is driven primarily by the \Btag reconstruction efficiency ($\sim$0.1\%) and, to a lesser degree, by the \SigLamb selection ($\sim60\%$). Systematic uncertainties in \SigEff arise from the \Btag correction factor $f_{B\bar{B}}$, the \SigLamb branching fraction, ${\rm BF}(\LambdaCtopkpi ) = \bfLamdaCtopkpi$~\cite{PhysRevD.110.030001}, and the limited statistical precision of the signal MC samples (1.2\%). An uncertainty of 0.6\% in the integrated luminosity is also propagated as a systematic uncertainty.

Signal yields are extracted as a function of \MPsi using a sliding-window method. Signal windows are defined as $\pm 5 \sigmaPsi$ centered around the nominal \MPsi value, with a step size equal to the mass resolution $\sigmaPsi$. For each mass hypothesis, the signal significance is determined using a profile likelihood method~\cite{Rolke:2004mj}. The background is treated as a Poisson process with the unknown mean estimated from the background simulation, while the efficiency is described by a Gaussian distribution with a standard deviation equal to the total systematic uncertainty.

\begin{figure}[ht]
\begin{center}
 \includegraphics[width=0.48\textwidth]{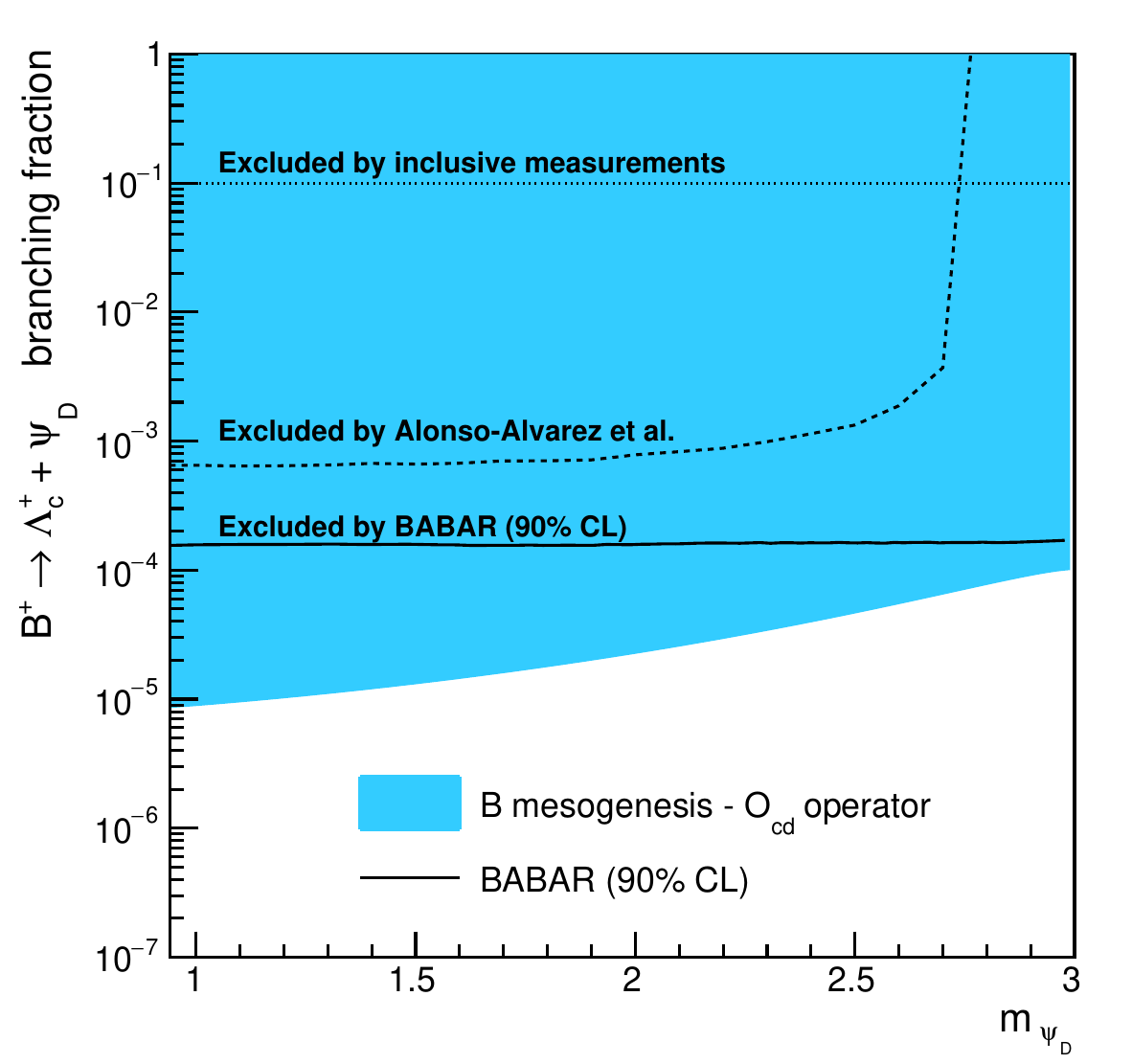}
\caption{The upper limits (90\% CL) on the \Sig branching fraction as a function of \MPsi. The colored area represents the region allowed by $B$ mesogenesis for the effective operator $\mathcal{O}_{cd}$ describing $b \to \SigPsi c d$ transitions. The curve labeled ``Excluded by Alonso-Alvarez {\it et al.}'' is a result from Ref.~\cite{Alonso-Alvarez:2021qfd} based on a reanalysis of data published by the ALEPH Collaboration~\cite{ALEPH:2000vvi}.}
\label{figBF}
\end{center}
\end{figure}

A total of 73 \MPsi hypotheses are tested in the range $0.94 < \MPsi < 2.99\GeV$. Since no signal is observed, the results are expressed as 90\% CL limits on the \Sig decay branching fraction. As shown in Fig~\ref{figBF}, these limits are at the level of $\SigBF < 1.6 - 1.7 \times 10^{-4}$, about an order of magnitude above the minimal branching fraction required for $B$ mesogenesis. These results improve upon indirect constraints by a factor of about five below $2.7 \GeV$, and three orders of magnitude above that threshold.

In conclusion, we report a search for a new baryonic dark sector particle $\psi_D$ in the process \Sig with a fully reconstructed \Btag meson. No significant signal is found, and constraints on the branching fraction $\SigBF < 1.6 - 1.7 \times 10^{-4}$ are derived at 90\% CL for $0.94 < m_{\psi_{D}} < 2.99 \GeV$. These results exclude a large fraction of the parameter space allowed for $B$ mesogenesis mediated by an effective operator $\mathcal{O}_{cd}$.

\section{Acknowledgments}
\label{sec:Acknowledgments}
We thank G. Elor and M. Escudero for useful discussions on the $B$ mesogenesis mechanism.
We are grateful for the extraordinary contributions of our \pep2\ colleagues in achieving the excellent luminosity and machine conditions that have made this work possible. The success of this project also relies critically on the expertise and dedication of the computing organizations that support \babar, including GridKa, UVic HEP-RC, CC-IN2P3, and CERN. The collaborating institutions wish to thank SLAC for its support and the kind hospitality extended to them. We also wish to acknowledge the important contributions of J.~Dorfan and our deceased colleagues E.~Gabathuler, W.~Innes, D.W.G.S.~Leith, A.~Onuchin, G.~Piredda, and R.~F.~Schwitters.

\bibliography{babar-pub-24001}
\end{document}